\documentclass[manuscript]{aastex}

\def\var{\hbox{AU\,Mon}}

\usepackage{graphicx}
 \usepackage{times}
  \usepackage{rotating}
  \usepackage{txfonts}

\slugcomment{}

\shorttitle{On the evolutionary stage of the interacting binary AU\,Mon}
\shortauthors{Mennickent, R.E.}

\begin{document}


\title{On the evolutionary stage of the interacting binary AU\,Mon}


\author{R.E. Mennickent\altaffilmark{1}}
\affil{Universidad de Concepci\'on, Departamento de Astronom\'{\i}a,
      Casilla 160-C, Concepci\'on, Chile}



 
\begin{abstract}
We present a study of the evolutionary stage of the interacting binary and Double Periodic Variable \var. 
A multi-parametric $\chi^{2}$ minimization is made between the observed parameters and those predicted  by the grid of non-conservative and conservative evolutionary models by Van Rensbergen et al., finding  the model that best represents the current
stellar and system parameters.  According to this model, the system started with initial masses 4 M$_{\odot}$ and 3.6 M$_{\odot}$ and orbital period 3.0 days,  196 million years ago, and at present undergoes a Case-B mass-exchange episode. 
This evolutionary stage  is consistent with the reported existence of a circumprimary accretion disk. However,  the  implied high mass transfer rate contrasts with the absence of significant orbital period change if the mass exchange is conservative. We show that this  can occur if the system has  recently entered in a non-conservative stage of mass transfer and the efficiency of mass and angular momentum loss satisfy certain conditions. 
\end{abstract}





\section{Introduction}

AU\,Mon (GCRV 4526, HD 50846, HIP 33237) is an interacting binary of orbital period 11.113 days consisting of a  mid-B
dwarf primary (or gainer) of temperature 17\,000 $\pm$ 2000 K accreting matter from a 
G-type giant secondary (or donor) of temperature 5750 $\pm$ 250 K (Peters 1994, Desmet et al. 2010).
The system and stellar parameters were recently determined by Djura\v{s}evi\'c et al. (2010) from the modeling of the CoRoT light curve assuming a semidetached configuration; 
they also find evidence for a geometrically thin and optically thick circumprimary accretion disk.

The system is characterized by a long photometric cycle of 416.9 days  (Desmet et al. 2010), and it is considered a member of the Galactic Double Periodic Variables, intermediate mass binary systems showing a long photometric cycle lasting about 33 times the orbital period (Mennickent et al. 2003, 2008, 2012a, Poleski et al. 2010). Peters (1994), based on its finding that the UV absorptions associated with the gas stream 
 (e.g. those observed in the line Si\,II\,1264 and  Si\,IV\,1402),  were weaker when the system was faint,  ascribed the long-period variation to changes in the mass transfer rate due to cyclic expansion and contraction of the secondary. 
This interpretation was challenged by Desmet et al. (2010)  who, 
based on  the long-term behavior of ground-based photometry, 
interpreted the long variability as attenuation of the total light by some variable circumbinary material. These authors also studied very accurate CoRoT space photometry finding 
suborbital frequencies of unknown origin.

Optical and UV spectra were used 
to model the temperature and structure of the accretion disk and the gas stream of \var\, by Atwood-Stone et al. (2012). That paper is a good source of references for the past literature of this interesting object. The disc found by these authors is about twice the size found by 
Djura\v{s}evi\'c et al. (2010); they argue that this is not inconsistent, since one model fits the optically thick disc component whereas the other is sensible to the optically thin disc envelope. It seems well established that 
the system shows stronger absorption lines during the low state (Peters 1994,  Atwood-Stone et al. 2012, Barr{\'{i}}a \& Mennickent 2011). The eclipse-mapping reconstruction of the accretion disc by Mimica \& Pavloski (2012) indicates a clumpy disc structure.

In this paper we study the evolutionary stage of AU\,Mon.
To our knowledge, no study of this type exist in the literature, hence
we expect to contribute to the understanding  of this interesting object. 
In Section 2 we present our methodology, in Section 3 our results and in Section 4 a discussion. We end in Section 5 with a summary of our main results.

\section{Methodology}

In this Section we compare the system parameters with those predicted by binary evolution models including epochs of non-conservative evolution. 
The stellar and system parameters given by Djura\v{s}evi\'c et al. (2010) were chosen since they provide the more realistic set of parameters  in the literature, incorporating in their model not only the stellar components but also the H$\alpha$ emitting  circumprimary accretion disc.   The parameters are given in Table 1; in this paper we use subscript h and c for the hot and cool star, respectively. 
In our study the orbital period of 11.1130374(1) days is taken from Desmet et al. (2010). We discuss the effect of the uncertainties associated to the system parameters in our methodology. 

We inspected the 561 conservative and non-conservative evolutionary tracks  by Van Rensbergen et 
al.  (2008), available at the Center de Donn\'ees Stellaires (CDS), looking for the best match for the system parameters found for AU\,Mon.
Models with strong and weak tidal interaction were studied, along with conservative models.
Following Mennickent et al. (2012a), a  multi-parametric fit was made with the  synthetic ($S_{i,j,k}$)  and observed ($O_{k}$) stellar parameters  mass, temperature, luminosity and radii, and the orbital period,  where $i$ (from 1 to 561) indicates the synthetic  model, $j$ the time $t_{j}$ and $k$  (from 1 to 9)  the  stellar or orbital parameter.
 Non-adjusted parameters were mass loss rate, Roche lobe radii, chemical composition, fraction of accreted mass lost by the system and age.
 For every synthetic model $i$ we calculated the quantity $\chi^{2}_{i,j}$ at every $t_{j}$ defined by

\begin{equation}
\chi^{2}_{i,j} \equiv (1/N) \Sigma_{k} w_{k}[(S_{i,j,k}-O_{k})/O_{k}]^{2},
\end{equation}
\noindent
where $N$ is the number of observations (9) and $w_{k}$ the statistical weight of the parameter $O_{k}$, calculated as

\begin{equation}
w_{k} = \sqrt{O_{k}/\epsilon(O_{k})},
\end{equation}
\noindent
where $\epsilon(O_{k})$ is the error associated to the observable $O_{k}$.
The model with the minimum $\chi^{2}$ corresponds to the model with the best evolutionary history of AU\,Mon. The absolute minimum $\chi^{2}_{min}$  identifies the age of the system along with the theoretical stellar and orbital parameters. The high accuracy of the orbital period dominates the search for the best solution in a single evolutionary track, but the others parameters play a role when comparing tracks corresponding to different initial stellar masses. 

\begin{table}
\centering
 \caption{The stellar parameters used in this paper (from Djura\v{s}evi\'c et al. 2010) to find the best evolutionary model for AU\,Mon. } 
 \begin{tabular}{@{}lclc@{}} \\
 \hline
 \hline
{\rm quantity}  &{\rm value} &  {\rm quantity} &{\rm value} \\
 \hline
$M_{c}$  &1.2 $\pm$ 0.2 M$_{\odot}$& $M_{h}$  &7.0 $\pm$ 0.3 M$_{\odot}$ \\
$T_{c}$ & 5754 $K$& $T_{h}$ & 15885 $K$\\
log $T_{c}$ &3.760 $K$  &log $T_{h}$ &4.201 $\pm$ 0.025   $K$\\
$L_{c}$ & 96.4 L$_{\odot}$ &$L_{h}$ & 1380.4 L$_{\odot}$ \\
log $L_{c}$ &1.984 $\pm$ 0.036 L$_{\odot}$ &log $L_{h}$ &3.140 $\pm$ 0.160 L$_{\odot}$\\
$R_{c}$ &10.1 $\pm$ 0.5  R$_{\odot}$ &$R_{h}$ &5.1 $\pm$ 0.5   R$_{\odot}$  \\
\hline
\end{tabular}
\end{table}

\begin{table}
\centering
 \caption{The parameters of the Van Rensbergen et 
al.  (2008) model that best fit the \var\,data of  Table 1. Parameters for the model with second lower $\chi^{2}$ are also given. The 
hydrogen and helium core mass fractions are given for the cool   ($X_{cc}$ and $Y_{cc}$)  and hot star  ($X_{ch}$ and $Y_{ch}$). Errors represent 
the full grid step at a given parameter.} 
 \begin{tabular}{@{}lcl@{}} \\
 \hline
 \hline
{\rm quantity}  &{\rm best model} &  {\rm 2nd model}  \\
 \hline
age & 1.96114(1)E8   yr  &2.2442(1)E8    yr                                                 \\
$M_{c}$  &1.533(5) M$_{\odot}$& 1.102(1)            M$_{\odot}$                    \\
$\dot{M_{c}}$ &-7.6(3)E-6   M$_{\odot}$ yr$^{-1}$ &-1.4(1)E-7 M$_{\odot}$ yr$^{-1}$\\   
$\dot{M_{h}}$ &7.5(1)E-6 M$_{\odot}$ yr$^{-1}$&1.4(1)E-7 M$_{\odot}$ yr$^{-1}$\\
$T_{c}$ &4966 $\pm$ 11 $K$ & 5358 $\pm$ 12 $K$ \\ 
log $T_{c}$ &3.696(1)   $K$ &3.729(1) $K$ \\
$L_{c}$ &66.1$^{+0.4}_{-0.5}$ L$_{\odot}$ &71.4$^{+0.4}_{-0.3}$ L$_{\odot}$\\
log $L_{c}$ &1.820(3) L$_{\odot}$ &1.854(2) L$_{\odot}$\\
$R_{c}$ &11.00(5)  R$_{\odot}$ &9.78(1)  R$_{\odot}$ \\
$X_{cc}$ &0.000(0) &0.000(0)  \\
$Y_{cc}$ & 0.980(0) &0.980(0)   \\   
$X_{ch}$ &0.324(0)&0.416(0) \\
$Y_{ch}$ &0.656(0) &  0.564(0) \\ 
period &11.1(1) d &11.12(3) d \\
$M_{h}$  &6.067(5) M$_{\odot}$ &5.298(1) M$_{\odot}$\\
$T_{h}$ & 16943 $\pm$ 39 $K$ &15740 $\pm$ 0 $K$\\
log $T_{h}$ &4.229(1)   $K$& 4.197(0) $K$ \\
$L_{h}$ &1914.3 $\pm$ 4.4  L$_{\odot}$ & 986.3 $\pm$ 2.3 L$_{\odot}$\\ 
log $L_{h}$ &3.282(1) L$_{\odot}$& 2.994(1) L$_{\odot}$\\
$R_{h}$ &5.082(0)   R$_{\odot}$&4.216(0) R$_{\odot}$\\
\hline
\end{tabular}
\end{table}

\section{Results}

We find the best model for the system with $\chi^{2}$ = 0.045 and another nearby model with $\chi^{2}$ = 0.050. Both solutions are clearly separated from the other models, having much lower  $\chi^{2}$  values, as illustrated in Fig.\,1. The two nearby points at the 2nd solution represent two subsequent time steps in the same model. Parameters for both solutions are given in Table 2. In order to test the stability of our result to the errors of the parameters given in Table\,1, we performed several trials moving the input parameters between the  values allowed by their uncertainties. 
We obtained in 86\% of the cases the aforementioned best solution, and in the remaining 14\%  the second solution.  Hence, our results are robust regarding the errors of the input parameters.
No other solution was found. This result might be interpreted as how much probable is the best solution regarding the second one.

The absolute $\chi^{2}$ minimum corresponds to the conservative 
model with initial masses of 4 M$_{\odot}$ and 3.6 M$_{\odot}$ and initial orbital period of 3.0  days.
 The best model gives a temperature of 16900 K for the B star, which is consistent with the most frequently quoted effective temperature for the B star, and better than the relatively low value given by Djura\v{s}evi\'c et al. (2010).
The corresponding evolutionary tracks for the primary and secondary stars are shown in Fig.\,2, along with the position for the best model for AU\,Mon and also the observational data. We observe a relatively good match for the gainer and donor  parameters. The best fit also indicates that  AU\,Mon is found inside a burst of mass transfer, the first one in the life of this binary (Fig.\,3). The donor is an inflated ($R_{c}$ = 11.0 R$_{\odot}$) and evolved 1.5 M$_{\odot}$ star with its core completely exhausted of hydrogen. According to the best model the system has now an age of 1.96 $\times$ 10$^{8}$ yr,  a mass transfer rate $\dot{M_{c}}$ =  -7.6 $\times$ 10$^{-6}$  M$_{\odot}$ yr$^{-1}$ and a mass ratio $q$ = 0.25.

The small difference between $\dot{M_h}$ and  $\dot{M_c}$ for this model in Table 2 is a purely  numerical effect due to the iterative process of calculation; the transferred mass equals the accreted mass in the long-term, i.e. according to the best model AU\,Mon  is in a conservative mass transfer stage. Moreover, this model is fully conservative at all epochs. 
The system is found 264,000 years after the beginning of the mass transfer. In this interaction episode the gainer has eventually accreted  2.47  M$_{\odot}$.


The second solution presented in Table 2 has   $\dot{M_{c}}$  = -1.4 x 10$^{-7}$ M$_{\odot}$/yr. In this case the system started with initial masses of 4 M$_{\odot}$ and 2.4 M$_{\odot}$ and orbital period 2.5 days; it is now in the 2nd episode of Roche-lobe contact during its lifetime and has a mass ratio $q$ = 0.35 (Fig.\,4). This figure is larger than the value $q$ = 0.17 reported by Desmet et al. (2010) and also larger than those found with the best solution, viz.\, $q$ = 0.25. 
The second solution gives an older system (2.24 $\times$ 10$^{8}$ yr)  and less massive and smaller stars than the best solution. 


\begin{figure}
\scalebox{1}[1]{\includegraphics[angle=0,width=8.5cm]{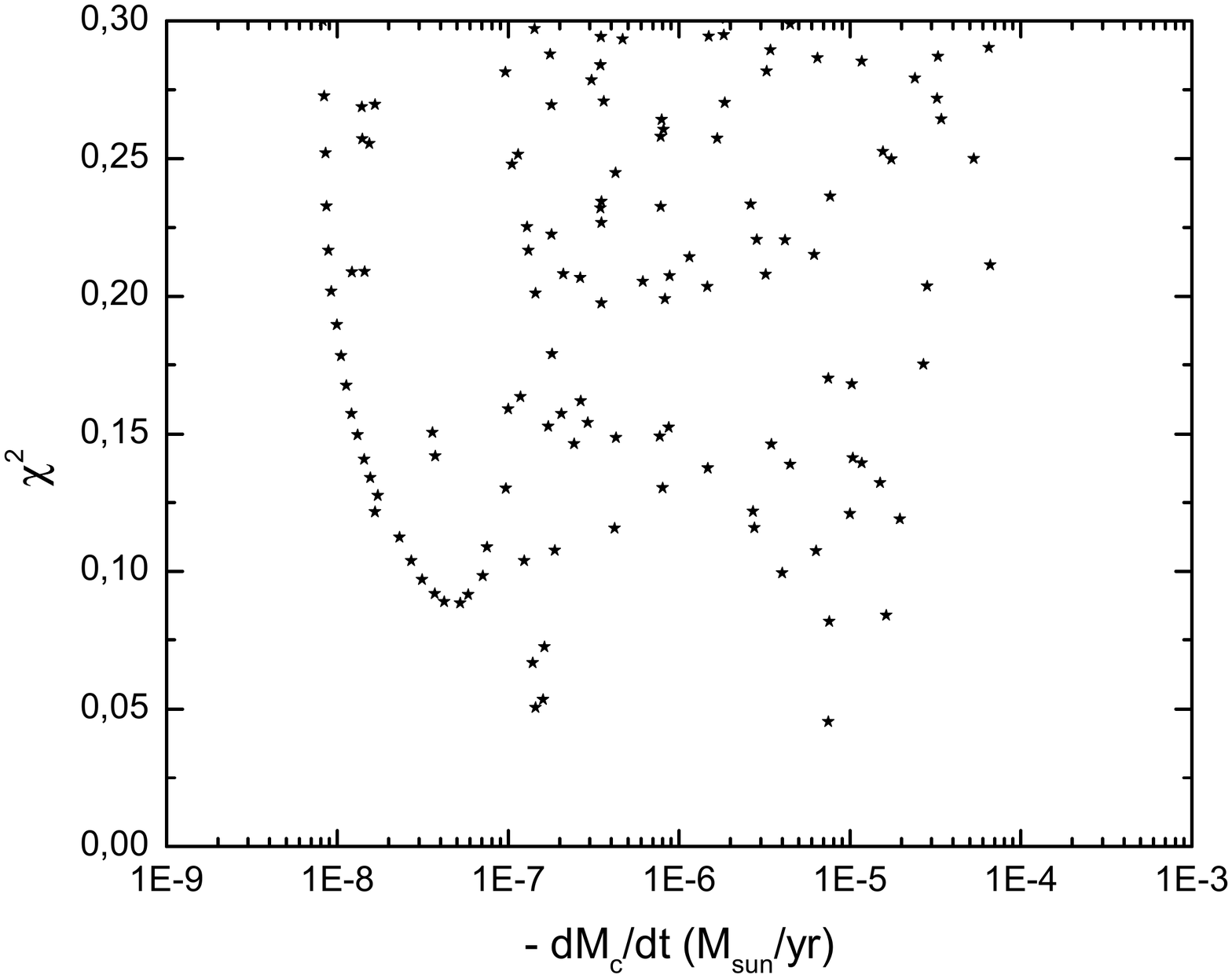}}
\caption{A zoom into the $\chi^{2}$ versus $dM_{h}/dt$ space showing the best possible solutions. The solutions corresponding to the minima with $dM_{h}/dt$ around 10$^{-7}$ and 10$^{-5}$ M$_{\odot}$/yr are discussed in this paper.}
  \label{x}
\end{figure}

\begin{figure}
\scalebox{1}[1]{\includegraphics[angle=0,width=16cm]{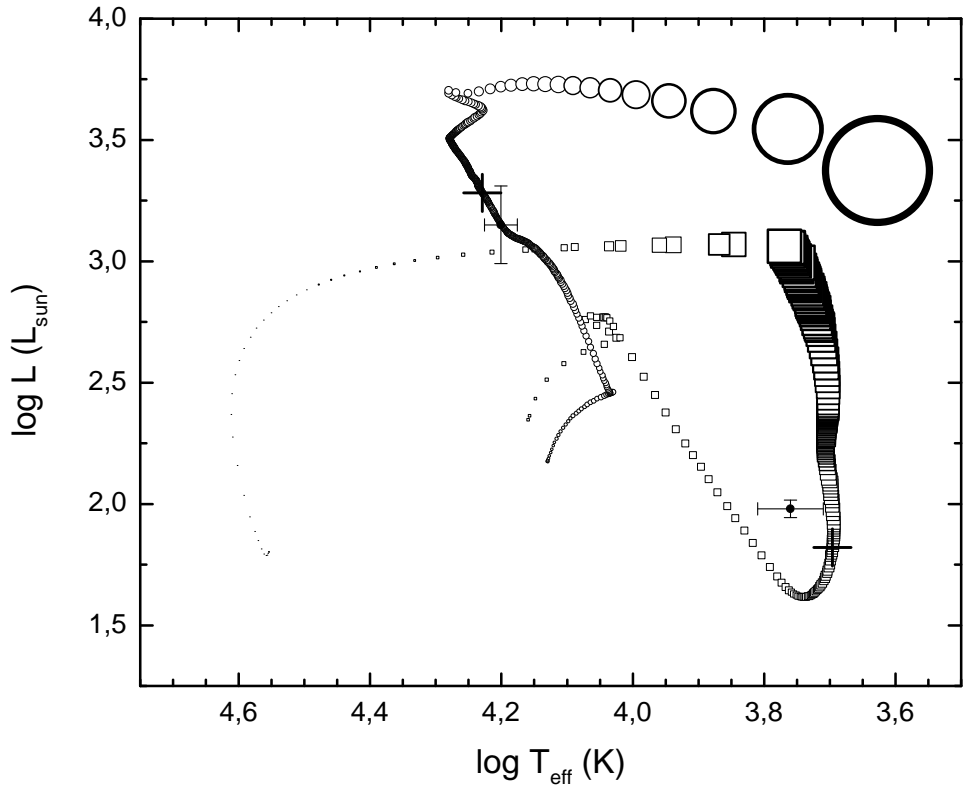}}
\caption{Evolutionary tracks for the  binary star model from Van Rensbergen et al. (2008) that best fit the data. Donor (right track) and gainer (left track) evolutionary paths are shown, 
along with the observations for AU\,Mon (with error bars, Djura\v{s}evi\'c et al. 2010). The best fit is reached at the time corresponding to the model indicated by large crosses, that is characterized in Table 2. Stellar 
sizes are proportional to the symbol size.  }
\label{lcs}
\end{figure}

\begin{figure}
\scalebox{1}[1]{\includegraphics[angle=0,width=8.5cm]{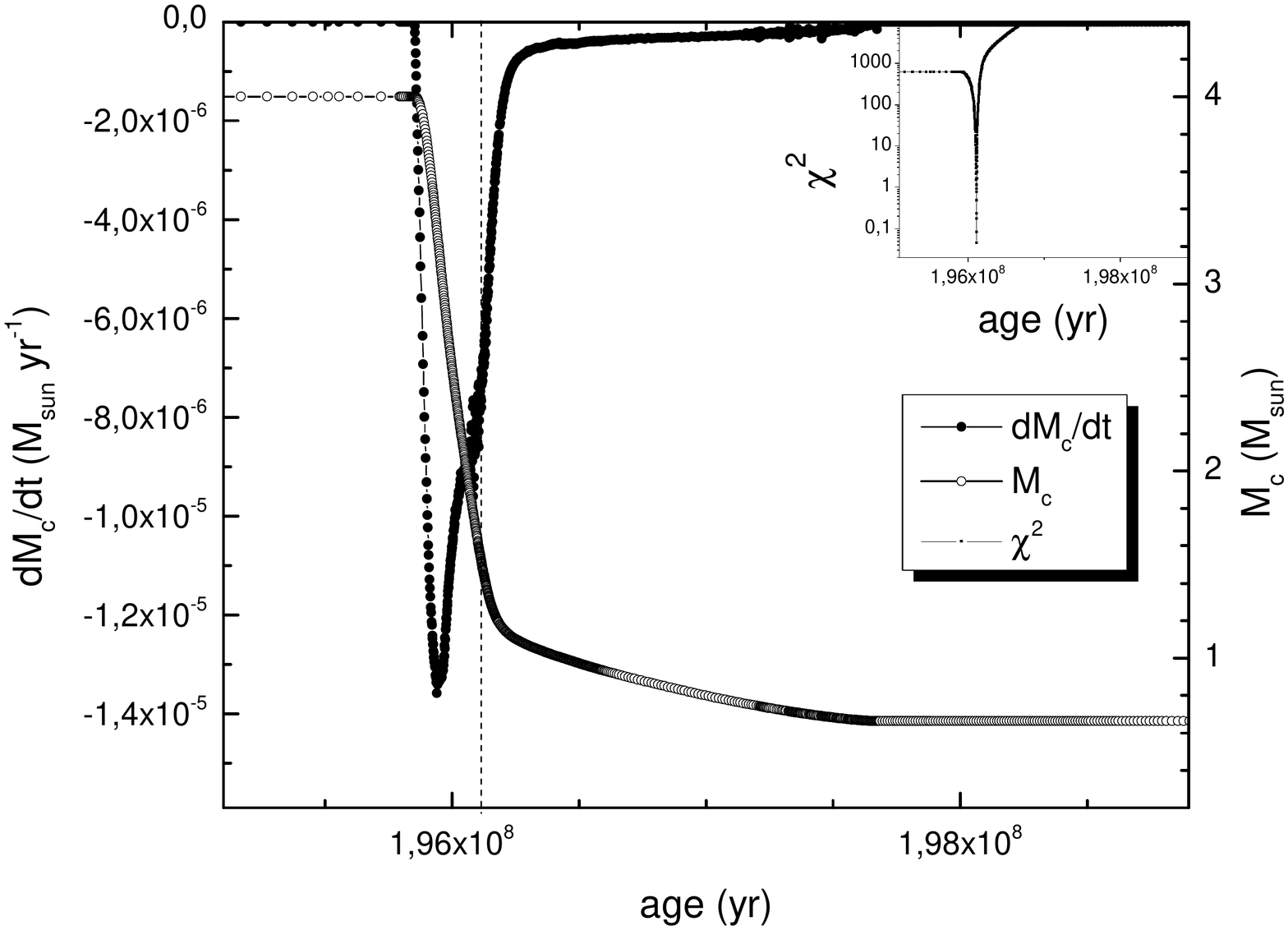}}
\caption{
$\dot{M_c}$ (upper curve) and $M_{c}$ for the best evolutionary model  in Table 2. The vertical dashed line  indicates the position for the best model. $\chi^{2}$ defined in Eq.\,2 is shown in the inset graph.
}
  \label{x}
\end{figure}

\begin{figure}
\scalebox{1}[1]{\includegraphics[angle=0,width=8.5cm]{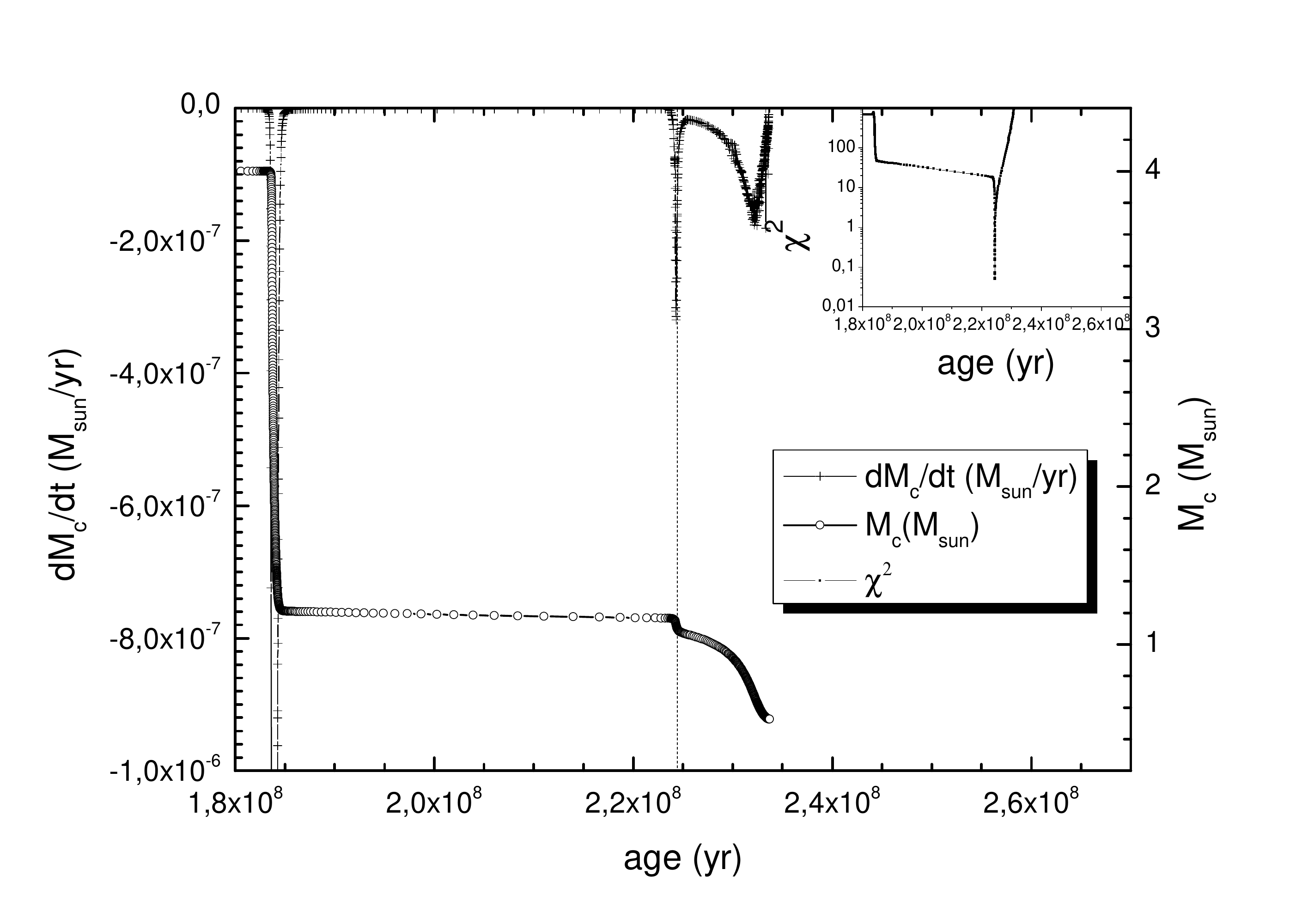}}
\caption{$\dot{M_c}$ (upper curve) and $M_{c}$ for the second evolutionary model in Table 2. The vertical dashed line  indicates the position for the best model. $\chi^{2}$ defined in Eq.\,2 is shown in the inset graph.}
  \label{x}
\end{figure}

\section{Discussion}

\subsection{On the mass transfer rate in AU\,Mon}

The mass transfer rate is one of the important parameters we get from the comparison with the grid of synthetic models. Our results can be compared with those found in the literature. For instance, Atwood-Stone et al. (2012) gives $\dot{M_{c}}$ = 2.4 $\times$ 10$^{-9}$  M$_{\odot}$ yr$^{-1}$, obtained from the physical parameters of the gas stream incorporated into the line profile models, although they quote that their estimate should be considered as a low limit. Actually, by lowering the stream temperature it is possible to obtain larger mass transfer rates for the same emissivity.
Peters (1987) reports  $\dot{M_{c}}$ $\geq$ 10$^{-12}$  M$_{\odot}$ yr$^{-1}$, measured from the redshifted absorption components in UV lines,  viz.  the resonance lines of Mg\,II, Al\,II, III and Si\,II, but she
indicates that this measurement  is model-dependent and complicated by  variability of underlying features. We notice also that the origin of UV lines  measured by Peters is not necessarily the stream or the accretion disc.
Peters (1987) mentions the similitude of $\dot{M_{c}}$ with the mass loss rate from the Be star due to its wind.
On the other hand, the very low
carbon abundance of $\log{N_C/N_{total}}$ = -5.4 determined from the strength of the C\,II\,4267 line in AU\,Mon suggests a mass transfer rate $\dot{M_{c}}$ $\sim$ 6 $\times$ 10$^{-6}$  M$_{\odot}$ yr$^{-1}$ (Ibano\v{g}lu et al. 2012). From the above we conclude that previous determinations of the mass transfer rate in AU\,Mon  are not necessarily in conflict with our values given in Table 2.

Based on data taken on a base time of 31.2 years, Desmet et al. (2010) give an uncertainty for the orbital period of 1 $\times$ 10$^{-7}$ d. This implies that a period change at constant rate, if present, should be less than 5.5 $\times$ 10$^{-4}$ s/yr to be consistent with the photometric time series. This is a notable observational constraint, since the relatively high $\dot{M_{c}}$ found in Section 3 should produce,  {\it  in the conservative case}, a constant orbital period change given by

\begin{equation}
\frac{\dot{P}}{P} = -\frac{3(1-q) \dot{M_{c}}}{M_{c}},
\end{equation}
\noindent (Huang 1963). For the AU\,Mon parameters this equation gives $\dot{P}/{P}$ =   1.58 $\times$ 10$^{-5}$ yr$^{-1}$,  or 15 seconds per year, which is definitively not observed. For the second solution the figure is 0.3 seconds per year. For these calculations we have used $M_{c}$ from Table 1 and  $\dot{M_{c}}$ from Table 2 but the results does not change qualitatively taking both parameters from the fit. 
The above result rises the question why the relatively high mass transfer rate does not produce a change in the observed orbital period. We propose a solution for this puzzling result.

Desmet et al (2010) proposed that the long cycles in AU\,Mon are due to attenuation of light by a variable circumbinary envelope.
This view was motivated by the constant shape of the orbital LC through the long cycle. However, we must note that a bipolar wind like that inferred for the Double Periodic Variable  V393\,Sco (Mennickent et al. 2012b) is also a possibility. In any case, photometric and spectroscopic evidence for mass and angular momentum  losses have been given for this system (Peters 1994, Desmet et al. 2010). These outflows could bring enough angular momentum to keep constant the orbital period.
In fact, Garrido et al. (2013) show that under some parameter combination, the effect of mass and angular momentum loss compensates that  of mass transfer, producing a constant orbital period, a fact already suggested by Mennickent et al. (2008). This can happen even for relatively high $\dot{M_{c}}$ DPV systems (e.g. OGLE05155332-6925581; Garrido et al. 2013) and could also be happening in AU\,Mon as we will show in the next section.

Alternative explanations for the apparent inconsistency between high mass transfer rate and constancy of the orbital period are: (i) that the models do not reproduce completely the physics of the system and (ii) that the density of initial parameters in the grid of binary star evolutionary models is not enough to adequately fit the object.  The first alternative is impossible to handle without extra (currently unavailable) theoretical mass loss prescriptions. We notice however that the view that AU\,Mon is the result of heavy mass loss from an initially much more massive binary conflicts with the absence of nebulosity in optical and infrared field images, for instance those  provided by the Wide field Infrared Survey Explorer in four bands centered at 3.4, 4.6, 12 and 22 $\mu$ (Wright
et al. 2010)\footnote{http://irsa.ipac.caltech.edu/Missions/wise.html}. 
In addition, the spectral energy distribution up to 22$\mu$ does not show evidence of a circumbinary dust shell, but it is dominated by the two stellar components  (Rosales \& Mennickent 2012). The second alternative could produce a mismatch between system ages, being the true system older or younger, still inside the mass transfer burst but with lower $\dot{M_{c}}$. 
The rapid changes in $\dot{M_{c}}$ during the mass transfer event make it quite sensitive to the age of the system.
However,  we notice that the same methodology (and models grid) applied to $\beta$ Lyrae has resulted in system parameters in close agreement with those previously calculated by other authors (Mennickent \& Djura{\v s}evi{\'c} 2013). This result gives support to the methodology used in this paper, and drives our attention to the possible existence of outflows carrying out mass and angular momentum from the system, in such a way of keeping constant the orbital period. This possibility is explored in the next subsection.


\subsection{On mass flows and angular momentum loss }

We have pointed  out the inconsistency between the relatively high mass transfer rate and the constancy of the orbital period in the 
conservative regime. The situation is completely different for the non-conservative case.

The models by Van Rensbergen et al. (2008) parametrize mass and angular momentum loss from the system through the parameters $\beta$ and $\eta$, respectively.  The mass loss is driven by radiation pressure from a hot spot located in the stream impact region on the stellar surface or accretion disc edge. The mass loss extracts angular momentum from the system and in this particular model only from the gainer. According to the models of Van Rensbergen et al. (2008), particular pairs of  the parameters $\beta$ and $\eta$ should result in a constant orbital period. Now we follow closely  the reasoning given by Garrido et al. (2013).

The mass and angular momentum loss from a system during mass transfer can be described with the $\left(\beta,\eta\right)$-mechanism (see Rappaport et al. 1983). Here the mass transfer efficiency is described by
\begin{equation}
\beta = \left|\frac{\dot{M_h}}{\dot{M_c}}\right|
\end{equation}
i.e. the fraction of mass lost by the donor 
accreted by the gainer. 
The angular momentum loss $\dot{J}$ for a given mass loss $\dot{M}$ ($\equiv  \dot{M}_{c}~(1-\beta)$) is determined by $\eta$, according to

\begin{equation}
\dot{J}=\sqrt{\eta}~\dot{M}_{c}~(1-\beta)~\frac{2~\pi~a^{2}}{P},
\end{equation}
being  $a$ the orbital separation and $P$ the orbital period. It can be shown (see e.g. Podsiadlowski et al. 1992) that whenever $\beta$ and $\eta$ remain constant throughout the considered period of time  (with indices i for initial and f for final), the resulting period variation is given by \\
\\
for $0 < \beta < 1$:

\begin{equation}
\frac{P_\mathrm{f}}{P_\mathrm{i}} = \left(\frac{M_{c\mathrm{f}}+M_{h\mathrm{f}}}{M_{c\mathrm{i}}+M_{h\mathrm{i}}}\right)\left(\frac{M_{c\mathrm{f}}}{M_{c\mathrm{i}}}\right)^{3\left[\sqrt{\eta}\,\left(1-\beta\right)-1\right]}\left(\frac{M_{h\mathrm{f}}}{M_{h\mathrm{i}}}\right)^{-3\left[\sqrt{\eta}\frac{1-\beta}{\beta}+1\right]},
\end{equation}
\vspace{0.3cm}
for $\beta = 0$:

\begin{equation}
\frac{P_\mathrm{f}}{P_\mathrm{i}} = \left(\frac{M_{c\mathrm{f}}+M_{h\mathrm{f}}}{M_{c\mathrm{i}}+M_{h\mathrm{i}}}\right)\left(\frac{M_{c\mathrm{f}}}{M_{c\mathrm{i}}}\right)^{3\left(\sqrt{\eta}-1\right)}\mathrm{e}^{3\sqrt{\eta}\,\left(\frac{M_{c\mathrm{f}}-M_{c\mathrm{i}}}{M_{h\mathrm{i}}}\right)}.
\end{equation}

If one assumes that mass is lost with the specific orbital angular momentum of the gainer, it can be shown that this yields

\begin{equation}
\eta = \left(\frac{M_{c}}{M_{c}+M_{h}}\right)^{4},
\end{equation}
resulting in $\eta\ll1$. On the other hand, if one assumes that matter is lost by the formation of a circumbinary disc 
a typical value is $\eta=2.3$ (Soberman et al. 1997).

We have calculated the possible $\beta$-$\eta$ combinations allowing the period remains stable between the observational boundaries established during 31.2 years. We find the same result for the two $\dot{M_{c}}$ listed in Table 2 (Fig.\,5).

 \begin{figure}
    \centering
    \includegraphics[width=0.5\textwidth, height=0.35\textwidth]{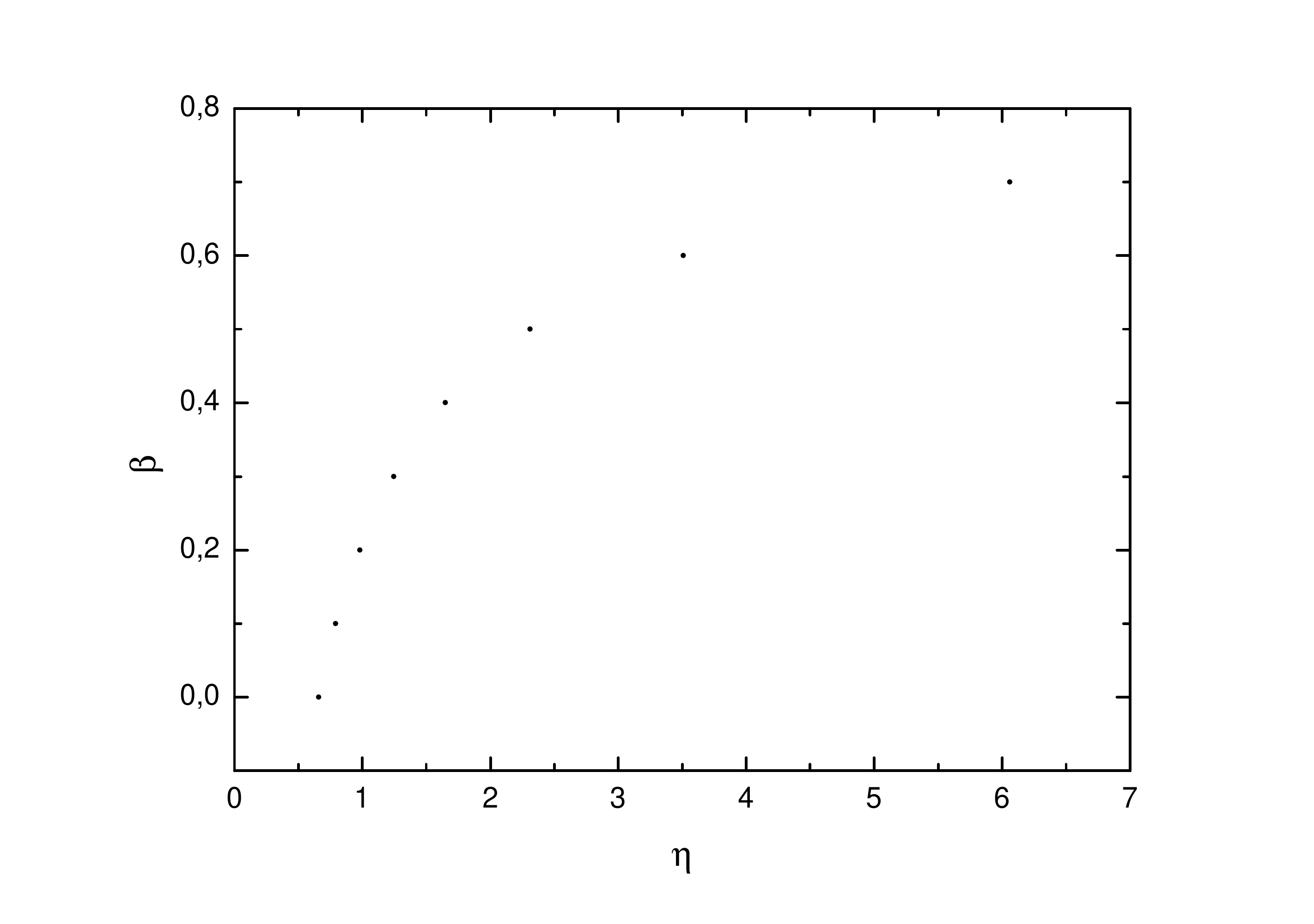}
    \caption{Some pairs of $\eta$ and $\beta$ allowing the constancy of the orbital period of AU\,Mon  according to Eq. (6) and (7).}
    \label{etabeta}
\end{figure}


Above the curve defined by the points the period increases too much; the case for large $\beta$ (little mass loss) and/or small $\eta$ (little angular momentum loss). Below the curve is where the period decreases too much; the case for small $\beta$ (much mass loss) and large $\eta$ (much angular momentum loss).

The assumption of specific gainer orbital angular momentum loss (the one used in the calculation of the models; Eq. 8) yields $\eta$ = 0.00046 for this system. This is too low  to fall into the constant period region; under this assumption, irrespective of $\beta$, the period will increase (much) more than allowed by the observations.
Based on this result we argue that another source of angular momentum loss is present in the system.
Notice that for $\eta \approx 2.3$,  representative for angular momentum loss from a circumbinary disc  (Soberman et al. 1997),
there is a mass loss ($\beta$ $\approx$  0.5) compatible with a constant period in this system. Interestingly, this is the same value needed in the Double Periodic Variable OGLE 05155332-6925581 for constancy of the orbital period 
(Garrido et al. 2013). 

If the above scenario is correct, AU\,Mon should be the second Double Periodic Variable with high mass transfer rate and constant orbital period after OGLE 05155332-6925581. Other DPVs as V\,393 Sco and DQ\,Vel also show stable orbital periods, but their mass transfer rates are not so large and the time baseline used for period determination is shorter than for AU\,Mon (Mennickent et al. 2012a, Barr\'{\i}a et al. 2013, 2014). It is possible that the same phenomenon causing the long cycle in DPVs keeps the orbital period constant. 

 If AU\,Mon is in a non-conservative stage, how we can trust in the best model described in Table 2 which is conservative?  
 The good agreement between observed parameters (stellar luminosities, radii, masses and temperatures), orbital period and 
semi-detached condition with those predicted by the best model strongly suggests that the departure from the conservative stage 
has not practical influence when comparing with a grid of 561 conservative and non-conservative evolutionary models. We think that the finding of the best model is not by chance but give us  an important  physical insight. The good match with the conservative model can be possible if AU\,Mon started  losing mass and angular momentum relatively recent in the binary life history. From the error in the stellar masses $\epsilon_{M}$ and the value for the mass transfer rate $\dot{M}$ we estimate this time of the order of  $\epsilon_{M}$/ $\dot{M}$ $\sim$  40\,000 years using  $\epsilon_{M}$ = 0.3 M$_{\odot}$. Within this time no appreciable departure from the conservative model is expected. In this case most of the binary star evolution has been conducted under a conservative regime even during the mass transfer episode lasting a tiny fraction of the binary age.  For the above considerations the parameters derived from the best model could be a good representation of the system at the current stage.

\section{Conclusions}

We have studied the evolutionary stage of the interacting binary and Double Periodic Variable AU\,Mon applying the methodology of a multi parametric $\chi^{2}$minimization between the observationally derived parameters and those of the grid of synthetic binary star evolutionary models by Van Rensbergen et al. (2008). A best model is found, followed by a close solution in the $\chi^{2}$ space. Parameters for both models are given in Table 2. 
It is still possible that the evolution of AU\,Mon follows other evolutionary route, but there is no way to inquire about this possibility, due to our current  limitation on the processes of mass and angular momentum loss in close interacting binaries. The  relevant physics known to day and a model for mass-loss  is included in the grid of models considered in this paper. Keeping this in mind,  our main results can be summarized as:\\

\begin{itemize}
\item We find AU\,Mon presently inside an episode of mass transfer with age 196 million years. This evolutionary stage is consistent with the reported existence of a circumprimary accretion disc.
\item The best model indicates that the system has a donor exhausted of hydrogen in its core, transferring mass at relatively high rate of 7.6 $\times$ 10$^{-6}$  M$_{\odot}$/yr. 
\item As the orbital period is relatively constant and $\dot{M_{c}}$ rather high, we speculate that outflows could be extracting angular momentum from the system, keeping constant the orbital period. This view is  consistent with reported observational evidence for  mass loss by Peters (1994) and the hypothesized circumbinary material  by Desmet et al. (2010). Furthermore, we show that under certain conditions of mass and angular momentum  losses, the system orbital period can be keep it constant even at  high mass transfer rate regimes.
\item The good agreement between the observed stellar and system parameters and those predicted by the best (conservative) model suggests that the non-conservative mass exchange is relatively recent, having the previous evolution being conducted mostly under a conservative regime.
\item We find that if matter is lost forming a circumbinary disc 
a mass loss rate about half the mass transfer rate ($\beta \approx 0.5$) should keep constant the orbital period. 
\end{itemize}

\acknowledgments

  We acknowledge the anonymous referee who helped to improve a first version of this paper.
 REM acknowledges support by VRID 214.016.001-1.0, Fondecyt grant 1110347 and the BASAL Centro de Astrofisica y Tecnologias Afines (CATA) PFB--06/2007. 
 REM also acknowledges Petr Harmanec for useful discussion on AU\,Mon during october 2012 in Prague (and june 2013 in Rodhes) and the hospitality of Pavel Koubsky
during a research stay in this city. REM acknowledges Nicky Mennekens and Jan Budai for useful discussions during the preparation of this paper.

\end{document}